\documentclass[prd, aps, superscriptaddress, preprintnumbers, floatfix, nofootinbib]{revtex4} %revtex4-2
\pdfoutput=1

\usepackage{amsfonts}
\usepackage{amsmath}
\usepackage{amssymb}
\usepackage{bm}
\usepackage{dcolumn}
\usepackage{graphicx}
\usepackage[latin1]{inputenc}
\usepackage{latexsym}
\usepackage{rotating}
\usepackage{hyperref}
\usepackage{graphicx}
\usepackage{xcolor}

\def\be{\begin{equation}}
\def\ee{\end{equation}}
\def\bea{\begin{eqnarray}}
\def\eea{\end{eqnarray}}

\begin{document}

\title{PERSPECTIVES ON THE DARK SECTOR}

\author{Robert Brandenberger}
\email{rhb@physics.mcgill.ca}
\affiliation{Department of Physics, McGill University, Montr\'{e}al, QC, H3A 2T8, Canada}

\date{\today}

%%%%%%%%%%%%%%%%%%%%%%%%%%%%%%%%%%%%%%%%%%%%%%%%%%%%%%%%%%%%%%%%%%%%%%%%%%%%%%%%%%%%%%%%%%%%%%

\begin{abstract}
 
I present some new perspectives on Dark Matter,  Dark Energy and the origin of structure in the Universe.  First, I argue that in order to understand the two latter issues, one needs to go beyond a standard point particle effective field theory analysis. Next, I review recent work attempting to construct a unified dark sector model from Heterotic superstring theory.  I finish by discussing a new research effort to obtain early Universe cosmology directly from a non-perturbative definition of superstring theory.
 
\end{abstract}
%%%%%%%%%%%%%%%%%%%%%%%%%%%%%%%%%%%%%%%%%%%%%%%%%%%%%%%%%%%%%%%%%%%%%%%%%%%%%%%%%%%%%%%%%%%%%%

\pacs{98.80.Cq}
\maketitle

\section{Introduction}

The nature of {\it Dark Matter} and of {\it Dark Energy} are two mysteries about the universe which were discussed in detail at this Blois workshop. A third mystery concerns the {\it origin of the observed structure}.  Each of these mysteries has a simple explanation.  For {\it Dark Energy}, the simple explanation is that it is a {\it cosmological constant}.  For a long time, the paradigm for {\it Dark Matter} has been the {\it WIMP} (weakly interacting dark matter particle),  and the current paradigm for the {\it origin of structure} is {\it cosmological inflation}.

However, these three simple explanations may all be wrong.  The {\it Trans-Planckian Censorship Conjecture} (TCC) \cite{BV} (see also \cite{BBLV}) implies that - at least at the level of an effective field theory description - Dark Energy cannot be a cosmological constant. Many talks at this workshop discussed the challenges which the WIMP paradigm faces, and suggested interesting alternative models.  What is less appreciated is that the same arguments which lead to the TCC impose tight constraints on inflationary cosmology, and indicate the the origin of structure might be very different.

In this summary talk I will focus on the two mysteries which were less discussed at the workshop, namely the {\it Dark Energy} and {\it origin of structure} mysteries, and indicate some new approaches.

The outline of this talk is as follows. I will first review the TCC and its implications for both early and late time cosmology.  Then, I will show that inflationary cosmology is not the only early universe scenario which is consistent with current data.  In Section 4 I will summarize an attempt \cite{BBF} to obtain a Unified Dark Sector model (a model describing both Dark Matter and Dark Energy) from superstring theory.  Based on the TCC and other considerations, it appears that we need to go beyond point particle effective field theory (EFT) in order to describe the very early universe. In Section 5 I will describe a recent reserach program \cite{us1} in which we start with a proposed non-perturbative definition of superstring theory, the BFSS matrix model \cite{BFSS},  and attempt to obtain an emerging metric space-time and early universe cosmology\cite{us2}.

 \section{Trans-Planckian Censorship}
 
 Let us consider the space-time diagram of an inflationary universe (see Fig. 1). Inflation can provide a causal mechanism for the origin of structure because scales which are being observed today (comparable to the current Hubble radius \footnote{The Hubble radius at time $t$ is defined as $H^{-1}(t)$, where $H$ is the Hubble expansion rate. The Hubble radius is relevant for the evolution of fluctuations. In terms of the canonical variables which describe them, fluctuations oscillate on sub-Hubble scales, but are frozen out and undergo squeezing when the wavelength is larger than the Hubble radius (see \cite{MFB} or \cite{RHBfluctsrev} for reviews of the theory of cosmological perturbations).} ) originate on sub-Hubble scales during the inflationary phase. However, as pointed out some time ago \cite{Jerome}, if the inflationary phase lasts only slightly longer than it has to last in order to explain the isotropy of the cosmic microwave background (CMB), then the initial wavelength of the fluctuations at the beginning of the inflationary phase was smaller than the Planck length (or string length, whichever is larger). Hence, it is clear that new physics must enter if we are to understand the generation of the fluctuations. Conversely, this argument implies that Planck-scale (or string scale) physics is testable in cosmological observations.
 
\begin{figure}
\centerline{\includegraphics[scale=0.5]{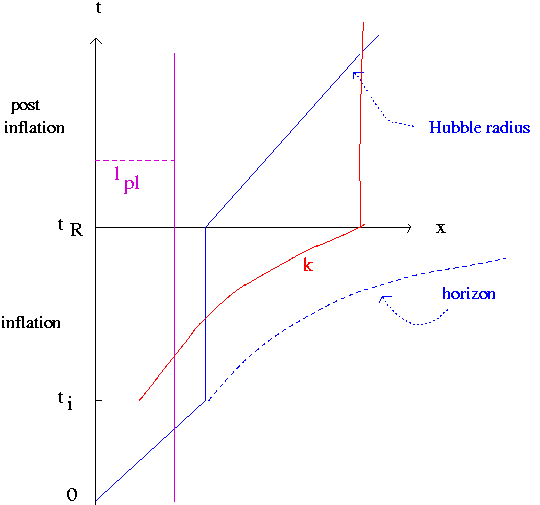}}
 \caption[]{Space-time sketch of an inflationary cosmology. The vertical axis is time, the horizontal axis corresponds to physical spatial distance. Inflation (here modelled as a phase of exponential expansion) begins at time $t_i$ and end at time $t_R$.  The solid blue line corresponds to the Hubble radius, the dashed blue line is the horizon, and the red curve denotes the physical length of a fluctuation mode (which has constant comoving length). The purple vertical line is the Planck length.}
\label{fig:radish}
\end{figure}

 Based on these considerations, Bedroya and Vafa posited \cite{BV} a new conjecture, the {\it Trans-Planckian Censorship Conjecture} (TCC), which states that in no cosmological scenario consistent with string theory (quantum gravity more generally), modes which are initially trans-Planckian can ever exit the Hubble horizon.  Working in terms of the usual metric of a spatially flat homogeneous and isotropic space-time
 \begin{equation}
 ds^2 \, = \, dt^2 - a(t)^2 d{\bf x}^2 
 \end{equation}
(where $t$ is physical time, ${\bf x}$ are the comoving spatial coordinates, and $a(t)$ is the cosmological scale factor) the TCC states that
 \begin{equation} \label{TCC}
 \frac{a(t_R)}{a(t_i)} l_{pl} \, < \, H^{-1}(t_R) \,
 \end{equation}
 for any initial time $t_i$ and final time $t_R$, and $l_{pl}$ is the Planck length. For a decelerating cosmology, this condition is automatically satisfied since the Hubble radius grows faster than $a(t)$, but for an accelerating cosmology like in the case of inflation the TCC leads to nontrivial constraints.
 
 There are various ways to justify the TCC (see \cite{RHB} for more detailed discussions). Firstly, the TCC can be justified in analogy with Penrose's black hole censorship hypothesis which states that, in any consistent theory of gravity, black hole singularities must be hidden from an external observer by a horizon.  This implies that black holes with charge greater than the mass (which are possible in the effective theory of General Relativity) cannot occur in the ultraviolet complete theory.  We propose to generalize this principle to cosmology according to the following conversion principle: position space $\rightarrow$ momentum space,  black hole singularity $\rightarrow$ space of trans-Planckian modes, black hole horizon $\rightarrow$ Hubble horizon. Then, the translated censorship hypothesis states that in no theory which is complete in the ultraviolet (UV) trans-Planckian modes can ever become super-Hubble. 
 
 As already discussed in \cite{Weiss} (and recently in \cite{Andy}), effective field theories (EFTs) in an expanding universe suffer from a unitarity problem. In effective field theories we expand all fields in Fourier modes and quantize each mode as a harmonic oscillator (with its corresponding ground state energy). To avoid the Planck catastrophy there needs to be an UV cutoff. While the modes which are being quantized are plane waves in comoving coordinates, the UV cutoff has to be at a fixed physical scale. To maintain this fixed UV scale in an expanding background, there needs to be a continuous production of Fourier modes: the Hilbert space of the EFT needs to be time-dependent. This is a violation of unitarity.  Unless the TCC is satisfied, the unitarity problem of the EFT can affect modes which can classicalize and become observationally measurable. This is the second justification for the TCC.
 
 In light of the above discussion, the cosmological constant (CC) problem arising from coupling quantum matter to gravity can be viewed as a consequence of an EFT analysis. If the correct quantum theory of space, time and matter is not based on an EFT, there is no reason to expect that the CC problem will arise - more about this later.
 
 The third motivation for the TCC comes from entropy considerations \cite{Omar}.  In an accelerating cosmological background, the entanglement entropy density between sub- and super-Hubble modes is an increasing function of time.  Demanding that the resulting entanglement entropy density remains smaller than the thermal entropy density at the beginning of the post-inflationary hot Big Bang phase leads to an upper bound on the duration of the accelerating phase which is consistent with the TCC.
 
 The TCC leads to severe constraints on inflationary cosmology. \cite{BBLV} . It implies an upper bound on the duration of iinflation which is given by (\ref{TCC}) with $t_i$ being the time when inflation starts and $t_R$ the time when it ends. On the other hand, it inflation is to provide a causal mechanism for the formation of strructure iin the universe, there is a lower bound on the duration of inflation. These bounds are only consistent (assuming exponential inflation) if the energy scale $\eta$ of inflation is small ($\eta < 3 \times 10^9 {\rm{GeV}}$), which requires tuning of the scalar field sector, and which leads to a negligible amplitude of primordial gravitational waves.
 
In conclusion, an EFT which does not satisfy the TCC is non-unitary and in violation of the second law of thermodynamics. This poses a severe challenge to EFT models of inflation. Luckily, there are other scenarios of early universe cosmology which can reproduce the current cosmological data, and which make interesting predictions for future data. Note that the TCC rules out a cosmological constant as the explanation for Dark Energy, since if Dark Energy were indeed a cosmological constant, modes which currently have a wavelength smaller than the Planck length would eventually exit the Hubble radius.
 
 \section{Scenarios for Early Universe Cosmology}
 
 In light of the problems in realizing cosmological iinflation, it is important to recall that inflation is not the only early universe scenario which is consistent with current cosmological observations.  In fact, ten years before the developnent of the inflationary scenario, Sunya'ev and Zel'dovich \cite{SZ} and Peebles and Yu \cite{PY} predicted acoustic oscillations in the angular power spectrum of the CMB, and baryon acoustic oscillations in the matter power spectrum, assuming only that some mechanism provided primordial adiabatic density fluctuations on length scales which were super-Hubble at times somewhat before the time of equal matter and radiation, and assuming that the spectrum of these fluctuations was approximately scale-invariant. Any early universe model which can produce such a spectrum of primordial fluctuations hence will be consistent with the data.
 
The question which was not answered in the pioneering work of \cite{SZ} and \cite{PY} is how to obtain an early universe model which yields such a spectrum.  Furthermore,  the reason for the overall isotropy of the CMB was still an open problem.  What are the key requirements for a successful early universe cosmology? Firstly,  in order to explain the isotropy of the CMB, the causal horizon (by which I mean the forward light cone of a point on an initial Cauchy surface) has to be much larger than the Hubble radius at the time of recombination, when the CMB is released.  Secondly, in order to obtain a causal theory for the origin of the primordial fluctuations on scales which are observed today, these scales have to originate inside of the Hubble horizon at some early time, since local physics can only affect the fluctuations on sub-Hubble scales. Thirdly, the resulting fluctuations must have a nearly scale-invariant spectrum.

The inflationary scenario \cite{Guth} is indeed the first proposed model based on causal physics in which the three above criteria can be realized.  In this context, the origin of fluctuations was worked out in \cite{Mukh} for cosmological perturbations and in \cite{Starob} for gravitational waves, assuming that the fluctuations start as quantum vacuum perturbations. However, there are other successful scenarios. In a bouncing cosmology (a cosmology which begins in a non-decelerating contracting phase, undergoes a bounce and emerges in the Standard Big Bang cosmology phase of expansion, the first two criteria are trivially satisfied since the horizon is infinite and the Hubble radius tends to infinity in the very far past and hence all modes start out sub-Hubble. In the same way that certain models of accelerated expansion lead to a scale-invariant spectrum, certain classes on bouncing models do as well \footnote{One example is the {\it matter bounce} scenario \cite{Fabio} in which the contracting phase is the mirror inverse of the expanding phase (but this model is unstable against the development of anisotropies \cite{Peter}), another is the Ekpyrotic scenario \cite{Ekpy} of super-slow contraction, and with an S-brane mediating the transition between contraction and expansion \cite{Ziwei}. Like in the case of inflation, it is assumed that fluctuations begin in a quantum vacuum state. See e.g. \cite{BounceRev} for a review of bouncing cosmologies.}.  

A third scenario in which the conditions for a successful early universe cosmology can be realized is the {\it emergent scenario} which is given by an initial emergent phase (which can be modelled as a quasi-static phase) which undergoes a phase transition into the Standard Big Bang phase of expansion. The resulting space-time diagram is sketched in Fig. 2. Since time is infinite in the past, the horizon is infinite. Since $H = 0$ deep in the emergent phase, the Hubble radius is also infinite. Hence,  the first two criteria for a successful early universe scenario are trivially satisfied.  It can be shown \cite{Nayeri} that if we start with thermal fluctuations in the emergent phase, and if these fluctuations have holographic scaling (e.g. the specific heat capacity in a volume of radius $R$ scaling as $R^2$), then the resulting density fluctuations and gravitational waves will obtain scale-invariant spectra. 

\begin{figure}
\centerline{\includegraphics[scale=0.35]{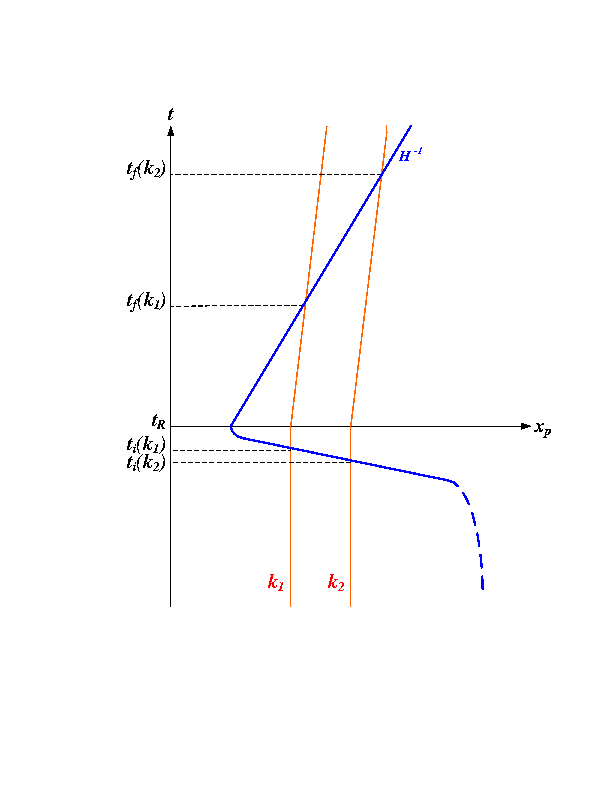}}
 \caption[]{Space-time sketch of an emergent cosmology. The vertical axis is time, the horizontal axis corresponds to physical spatial distance.  The time $t_R$ is the time of the transition between the quasi-static emergent phase and the Standard B ig Bang phase of expansion.  The solid blue line corresponds to the Hubble radius, the red lines labelled $k_1$ and $k_2$ indicate the physical length of fluctuation modes.  The time $t_i(k)$ is the time when the mode $k$ exits the Hubble radius.  }
\label{fig:pear}
\end{figure}

As long as the energy scale at the bounce point is smaller than the Planck scale, then the TCC is trivially satisfied in bouncing scenarios because no trans-Planckian modes ever cross the Hubble horizon.  Similarly, the TCC is trivially satisfied in emergent scenarios as long as the energy scale during the emergent phase is smaller than the Planck mass.
Thus, from the point of view of the TCC,  scenarios alternative to inflation appear to have advantages. Note, however, that the TCC does not apply to constructions of inflation which do not rely of EFTs (see e.g. \cite{Dvali} and \cite{Keshav} for two approaches). The bottom line of the discussion in this section is that we need to go beyond an EFT description of physics if we want to understand the early universe \footnote{In order to get a bounce, one needs to go beyond a standard EFT, and one cannot obtain an emergent phase sticking to EFTs coupled to Einstein gravity.}.

\section{Unified Dark Sector Model}
 
 For the moment, however, let us stick to EFT and ask whether one can obtain a unified dark sector model, a model which describes Dark Matter and Dark Energy in a unified way. Quintessence \cite{Ratra} is a prime candidate for Dark Energy.  In this scenario, the dark energy field is a scalar field with a potentail which is exponentially descreasing at large field values. As long as the duration of the Dark Energy phase is not too long, such a model can be consistent with the TCC, and also with the swampland constraints \cite{Lavinia}.  However, with scalar fields it is possible to model any given time evolution of the scale factor. Hence, to make a Quintessence model of Dark Energy appealing, one should try to get more than just Dark Energy out of it. In this spirit, we have recently proposed a toy model \cite{JF} \footnote{See also \cite{Stephon} for related work, and \cite{Elisa} for a unified dark sector model based on superfluid dark matter.} which unifies Dark Energy and Dark Matter. It is based on considering a canonically normalized complex scalar field $z = \sigma + i \theta$ with potential
\begin{equation} \label{toy}
V(\sigma, \theta) \, \simeq \, \bigl[ \Lambda + \frac{1}{2} \mu^4 {\rm{sin}}^2 (\theta / f) \bigr] e^{-2 \sigma / f} \,. ,
\end{equation}
where $\Lambda$ and $\mu$ are constants, and $f$ is a characteristic energy scale. Here, the field $\sigma$ is slowly rolling and yields Dark Energy, while $\theta$ oscillates about $\theta = 0$ and provides a Dark Matter candidate.

Initially, the energy density is dominated by regular radiation, $\theta$ is frozen in at the value $\theta = f$, and slow rolling of $\sigma$ can yield baryogenesis.  Once the energy density of radiation falls to a sufficiently low value,  the confining force on $\theta$ disappears and $\theta$ starts to oscillate.   This is the beginning of the dark matter phase.  Provided that $f$ is sufficiently large (comparable to the Planck mass $m_{pl}$), $\sigma$ will be rolling only very slowly, and the energy in the $\sigma$ component of the field will eventually dominate, yielding the onset of the Dark Energy phase.. Like in any Quintessence-type model, tuning of the parameters $\Lambda$ and $\mu$ is required in order to obtain the observed density of Dark Energy, and to solve the ``coincidence problem'', i.e. to explain why the onset of the Dark Energy phase is so close to the present time. Specifically,  requiring the onset of Dark Energy domination at the present time $t_0$ leads to the condition
\be
\Lambda e^{-2 \sigma(t_0) / f} \, \sim \, T_0^4 z_{eq} \, ,
\ee
where $T_0$ is the current temperature of the CMB and $z_{eq}$ is the redshift at the time of equal matter and radiation, while to reproduce the observed density of Dark Matter requires
\be
\mu^4 \frac{{\cal{A}}^2(T_0)}{f^2} e^{- 2 \varphi(t_0) / f} \, \sim \, T_0^4 z_{eq} \, \nonumber
\ee
where ${\cal{A}}$ is the amplitude of oscillation of $\theta$.

An interesting question is whether such a model can be realized starting from superstring theory. There are two major challenges: firstly how to generate the low energy scale of Dark Energy from a theory whose characteristic energy scale, the string scale, is close to the Planck scale,  and secondly how to generated a sufficiently flat potential to yield accelerated expansion.  These are the questions we recently tackled in \cite{BBF}. It turns out that it is easy to generate the low scale of Dark Energy, while it is much less natural to obtain a sufficiently flat potential, in line with what one would expect from the ``swampland'' conditions \cite{swamp}, conditions which all EFTs emerging from superstring theory appear to need to obey.  

Specifically,  we consider the model-independent axio-dilaton superfield $S \, = \, e^{- \Phi} + i a $ and the volume modulus field $T = e^{\Psi} + i b$.  Here, $\Phi$ is the ten-dimensional dilaton field and $a$ is its associated axion, $\Psi$ yields the radius of the extra-dimensional space, and $b$ is its associated axion (which we have set to zero in our analysis). The model-independent axion $a$ naturally yields Dark Matter, while $\Psi$ is a candidate for Dark Energy.  The first step in the analysis is to dimensionally reduce the 10-d effective action for heterotic superstring theory to four dimensions. In order to be consistent with late time cosmology,  the four dimensional dilaton $\phi$ needs to be stabilized. This can be achieved via the standard tool of gaugino condensation. Gaugino condensation then induces a non-vanishing superpotential $W$ for $S$
\be
W \, = W_0 - A e^{-a_0 S} \, ,
\ee
where $W_0, A$ and $a_0$ are constants. Making use of the standard Kaehler potential $K$ for $S$ and $T$, we can then derive the potential for $\phi$, $a$ and $\sigma$, where $\sigma$ is the canonically normalized field related to $\Psi$ via
 \be
\Psi \, = \, \sqrt{\frac{2}{3}} \frac{\sigma}{m_{pl}} \, .
\ee
The resulting potential is
 \bea \label{pot0}
    V(\phi, \sigma, a) &=& \frac{e^{-\sqrt{6}(\sigma/m_{pl})}}{8}e^{-\sqrt{2}(\phi/m_{pl})} \kappa^2 
    \Big[ A^2 e^{-2a_0e^{-\sqrt{2}(\phi/m_{pl})}}\Big(a_0  \nonumber \\
    & &+ \frac{1}{2}e^{\sqrt{2}(\phi/m_{pl})}\Big)^2 
    + \frac{W_0^2}{4}e^{2\sqrt{2}(\phi/m_{pl})} \\
    & &- AW_0\Big(a_0 +\frac{1}{2}e^{\sqrt{2}(\phi/m_{pl})}\Big)e^{\sqrt{2}(\phi/m_{pl})}e^{-a_0e^{-\sqrt{2}(\phi/m_{pl})}}\cos(a_0 a)\Big] \,. \nonumber
\eea
The last line corresponds to the axion term in the toy model potential (\ref{toy}). The first term in (\ref{toy}) results from integrating out fluctuations of the axio-dilaton field.  From the form of the potential it follows that only mild tuning on the initial value of $S$ is required to obtain the energy scale of Dark Energy. On the other hand, in order to obtain a sufficiently flat potential, we must by hand introduce an anisotropy of the extra-dimensional space and only allow one of the extra dimensions to have a time-dependent radius. Note that the Dark Matter which emerges is ultralight (for details the reader is referred to \cite{BBF}).

 \section{Emergent Cosmology from Matrix Theory}

Let us now turn to the very early universe. The analysis in the previous section was based on an EFT.  But we argued in earlier sections that a true understanding of the very early universe will require going beyond point-particle-based EFT. Here we summarize a recent approach \cite{us1} \footnote{See also \cite{RBrev} for a recent review.} to develop a new quantum picture of space, time, matter and early universe cosmology based on the BFSS model \cite{BFSS}, a model which has been suggested to yield a non-perturbative definition of superstring theory \footnote{See \cite{Ydri} and \cite{Taylor} for reviews of matrix models for superstring theory.}.

The BFSS model is a matrix model given by the Lagrangian
\begin{equation}
L \, = \, \frac{1}{2 g^2} \bigl[ {\rm Tr} \bigl( \frac{1}{2} (D_t X_i)^2 - \frac{1}{4} [X_i, X_j]^2 \bigr) \bigr] 
\end{equation}
where $X^i,  i = 1, \cdots 9$ are nine $N \times N$ Hermitean matrices, $g$ is a coupling constant, and $D$ is a matrix covariant derivative which is given by the partial derivative with respect to time plus a commutator with a tenth matrix $A_0$.  There are also fermionic matrices to render the model supersymmetric, but these will play little role in the following. Indices are raised with the Minkowski $\eta_{\mu \nu}$ tensor. In the limit $N \rightarrow \infty$ with $\lambda \equiv g^2 N $ held fixed one obtains a non-perturbative definition of superstring theory.

Our proposal \cite{us1} is to consider this matrix model in a high temperature state. In this case, the action of the model is dominated by the zero modes. We expand the matrices in Euclidean time in terms of Matsubara frequencies
\begin{equation}
X_i(t) \, = \, \sum_n X_i^n e^{2 \pi i T t} \, ,
\end{equation}
where $T$ is the temperature. If we rescale the zero modes of the matrices as
\begin{equation}
A_i \, \equiv \, T^{-1/4} X_i^0  \, ,
\end{equation}
 then the leading term (in a power series in $1/T$) of the bosonic part of the action  is given by the bosonic part of the IKKT action \cite{IKKT} which is
  \begin{equation} \label{IKKTaction}
S \, = \,  -\frac{1}{g^2} {\rm{Tr}} \bigl( \frac{1}{4} [A^a, A^b] [A_a,A_b]  \bigr) \,, .
\end{equation}
The partition function of the bosonic IKKT model is given by
\begin{equation}
Z \, = \, \int dA  e^{iS} 
\end{equation}
where the standard Haar measure is assumed. This is a quantum mechanical model which is not based on quantizing harmonic oscillator modes, and hence there is no reason to expect the usual cosmological constant problem to arise. 

Since the matrices are Hermitean, we can work in the basis in which $A_0$ is diagonal. The diagonal elements are ordered according to their increasing value, and yield {\it emergent time}. Numerical studies of the IKKT matrix model \cite{Nish1} indicate that
\be
\frac{1}{N} \bigl< {\rm Tr} A_0^2 \bigr> \, \sim \, \kappa N \, ,
\ee
where $\kappa$ is a constant., and the pointed brackets denote expectation values. This implies that the maximal temporal eigenvalue $t_{max}$ scales as $t_{max} \sim \sqrt{N}$ and hence the distance $\Delta t$ between the discrete time steps scales as $\Delta t \sim \frac{1}{\sqrt{N}}$. Hence, in the large $N$ limit, the emergent time becomes infinite and continuous. Since the temporal eigenvalues are also symmetric about $t = 0$, the emergent time is infinite both in the past and future.  Hence, there will be no Big Bang singularity.

\begin{figure}
\centerline{\includegraphics[scale=0.35]{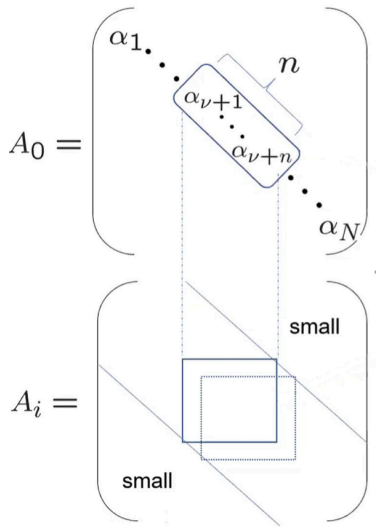}}
 \caption[]{Definition of the spatial submatrices ${\tilde{A_i}}(n_i , t)$. For a fixed time $t$, we consider $n_i \times n_i$ submatrices of the full spatial matrices $A_i$ which are centered a distance $t$ ``down the diagonal''.}
\label{fig:orange}
\end{figure}

Focusing on the spatial matrices $A_i$, for any time $t$ we can consider the $n_i \times n_i$ submatrices ${\tilde{A_i}}(n_i , t)$ centered at the position $t$ of the diagonal (see Fig. 3).  The expectation values of the squares of these submatices can be defined \cite{Nish2} as the {\it extent of space parameters} $x_i(n_i, t)$
\be
x_i(n_i, t)^2 \, \equiv \, \left\langle \frac{1}{n_i} {\rm{Tr}} ({\bar{A_i}})(n_i, t))^2 \right\rangle \, .
\ee
Numerical studies of the IKKT model have shown that the $SO(9)$ symmetry of the action is broken to $SO(3) \times SO(6)$, with exactly three of the extent of space parameters becoming large \cite{Nish2}.  The result can be verified by making use of Gaussian expansion calculations. This is the same symmetry breaking pattern observed in {\it String Gas Cosmology} \cite{BV2}. In the case of String Gas Cosmology, the reason why only three out of the nine dimensions of space can expand is that string winding modes (which are present in the initial thermal state) cannot intersect and decay in more than three large spatial dimensions. We expect that a similar mechanism is at work in matrix theory. We have recently \cite{us3} shown that the $SO(9)$ symmetry of the Lagrangian is also broken in the BFSS matrix model.

The off-diagonal elements of the spatial submatrices satisfy the following properties:
\be \label{prop1}
\sum_i \bigl< |A_i|^2_{ab} \bigr> \,\,\,   {\rm{decay \,\,\,when}} \,\,\, |a - b| > n_c
\ee
with $n_c \sim \sqrt{N}$, and
\be \label{prop2}
\sum_i \bigl< |A_i|^2_{ab} \bigr> \sim  \, {\rm{constant \,\,\, when}} \,\,\, |a - b| < n_c \, .
\ee
The first property can be seen directly from the action by means of the Riemann-Lebesgue Lemma, the second property has been observed numerically \cite{recent}.
  
Our proposal \cite{us2} to obtain an emergent metric space from the BFSS matrix model is to identify the $n_i$ as comoving spatial coordinates, and
\be
l_{phys, i}^2(n_i, t) \, \equiv \, \left\langle {\rm{Tr}} ({\bar{A_i}})(n_i, t))^2 \right\rangle \, 
\ee
as the (square of the) length of the curve along the $i$ axis from $n_i  = 0$ to $n_i$.  Then, the spatial metric elements $g_{ii}(n_i, t)$ can be obtained as
\begin{equation}
g_{ii}(n_i, t)^{1/2} \, = \, \frac{d}{dn_i} l_{phys, i}(n,_i t) \, .
\end{equation} 
Making use of the property (\ref{prop2}) it follows that this metric element is independent of $n_i$.  Assuming the $SO(3)$ symmetry we obtain the full spatial metric
\begin{equation}
g_{ij}({\bf{n}}, t) \, = \, {\cal{A}}(t) \delta_{ij} \,\,\,   i = 1, 2, 3 
\end{equation}
which corresponds to a homogeneous, isotropic and spatially flat space of infinite extent.
Hence, we do not require a period of inflation to solve the horizon and flatness problems of Standard Big Bang cosmology (note that the entire space inherits the same temperature from the matrix model).

At this point we have given a prescription for obtaining emergent metric space-time from the zero modes of the BFSS matrices, assuming that we are considering the BFSS model in a high temperature state. Since the state is thermal, there will be thermal fluctuations, and these fluctuations will lead to cosmological fluctuations and gravitational waves.

Assuming the validity if the linearized Einstein equations in the far infrared, the cosmological fluctuations $\Phi$ and gravitational waves $h$ are given by the expectation values of the energy-momentum tensor perturbations 
\bea
\langle|{\Phi}(k)|^2\rangle \, &=& \, 16 \pi^2 G^2 
k^{-4} \langle\delta T^0{}_0(k) \delta T^0{}_0(k)\rangle \, \\
\langle|{h}(k)|^2\rangle \, &=& \, 16 \pi^2 G^2 
k^{-4} \langle\delta T^i{}_j(k) \delta T^i{}_j(k)\rangle \,\,\, i \neq j \,  . \nonumber 
\eea
Here, $\Phi$ is the amplitude of the scalar metric fluctuation, and $h$ is the amplitude of a gravitational wave mode. 

In a thermal state, the correlation functions of the energy-momentum tensor $T_{\mu \nu}$ are given by partial derivatives of the finite temperature partition function. Following the same formalism which was used in \cite{Nayeri} in the context of String Gas Cosmology, we find \cite{us1} a scale-invariant spectrum of cosmological fluctuations with a Poisson contribution in the UV, and a scale-invariant spectrum of gravitational waves. Hence, the emergent early universe scenario which emerges from the BFSS matrix model is in agreement with current cosmological data. In analogy with what happens in String Gas Cosmology, we expect to obtain a slight red tilt of the spectrum of cosmological perturbations, and a slight blue tilt of the tensor modes. However, to compute these tilts we need an understanding of the phase transition from the emergent phase to the late time phase when three spatial dimensions are expanding.

At this stage, it appears that the BFSS matrix model can provide an emergent metric space-time which is infinite in both temporal and spatial directions, which is continuous and spatially flat.  Starting from a high temperature state of this matrix model, we obtain precisely three large spatial dimensions,  and the thermal fluctuations in this state lead to cosmological perturbations and gravitational waves with spectra in agreement with current observations.  The quantum mechanical matrix model is not based on quantizing harmonic oscillator modes, and hence there is no reason to expect a cosmological constant to arise. In fact, solving the late time classical matrix equations (for the bosonic sector) leads to a cosmological scale factor increasing at $t^{1/2}$, like in a phase of radiation \cite{late}. 

Our matrix cosmology research program is at the very beginning. It will be important to include the fermionic sector (which will provide the usual matter, and quite likely also the Dark Matter), to identify localized excitations which in the low energy limit become particle states, and to compute their interactions. This should lead to a low energy effective action.  A important question is whether the local gravitational interactions are given by the emergent cosmological metric, or by a different effective metric \footnote{I thank H. Steinacker for stressing to us this issue.}. Note that there are other approaches to obtaining a metric space-time and an early universe cosmology from the IKKT matrix model, in particular the approach of \cite{Stein1} which yields interesting novel cosmological solutions \cite{Stein2}. For an alternative non-perturbative approach to obtaining an emergent phase from superstring theory see also \cite{Vafa}.

\section{Conclusions}
 
 The first important message from this lecture is that {\it inflationary cosmology} is not the only early universe scenario which is consistent with current observations. In fact, in light of conceptual problems such as the TCC, alternatives such as the {\it emergent universe scenario} appear more promising.  The second message is that if we want to understand the early universe (and possibly also the very late universe) we need to go beyond point particle effective field theory.  After reviewing a proposal to obtain a unified dark sector model from Heterotic superstring theory,  I introduced a recent approach to obtain metric space-time and an emergent early universe cosmology from the BFSS matrix model, a proposed non-perturbative definition of superstring theory.
   
\section*{Acknowledgments}

I wish to thank my collaborators, in particular H. Bernardo,  S. Brahma, J. Fr\"ohlich and S.Laliberte, for their contributions to the work summarized here (errors in presentation are obviously mine).  I also than B. Ydri for permission to use Fig. 3 \cite{Ydri}.This work has been supported by NSERC, and by funds from the Canada Research Chair program. I am grateful for hospitality by and support from the Pauli Institute at the ETH Zuerich.

\end{document}